\documentclass[10pt,a4paper,twoside]{article}
\usepackage{epsfig}
\usepackage{baltlat6}
\usepackage{array}
\usepackage{here}
\begin{document}
\ \ \vspace{0.5mm} \setcounter{page}{1} \vspace{8mm}

\titlehead{Baltic Astronomy, vol.\,}

\titleb{THE INFLUENCE OF SOLAR SPECTRAL LINES ON\\ELECTRON CONCENTRATION IN\\ TERRESTRIAL IONOSPHERE}

\begin{authorl}
\authorb{A. Nina}{1},
\authorb{V. {\v C}ade{\v z}}{2},
\authorb{V. Sre{\'c}kovi{\'c}}{1} and
\authorb{D. {\v S}uli{\'c}}{3}
\end{authorl}

\begin{addressl}
\addressb{1}{Institute of Physics,
University of Belgrade,\\  Pregrevica 118, Zemun, 11080 Belgrade, Serbia; sandrast@ipb.ac.rs}
\addressb{2}{Astronomical Observatory,\\ Volgina 7, 11060 Belgrade, Serbia; vcadez@aob.rs}
\addressb{3}{Faculty of Ecology and Environmental Protection, \\Union - Nikola Tesla University,
\\ Cara Du\v sana 62, 11000 Belgrade, Serbia; dsulic@ipb.ac.rs}
\end{addressl}

\submitb{Received:           ; accepted:         }

\begin{summary} One of methods of detection and analyses of solar flares is by
observing time variations of certain solar spectral lines. During
solar flares, a raise of electron concentration occurs in Earth's
ionosphere which results into amplitude and phase variations of
recorded very low frequency (VLF) waves. We compared the data
obtained by the analysis of recorded VLF signals and line spectra
for different solar flares. In this paper we treated the DHO VLF
signal transmitted from Germany at the frequency of 23.4 kHz
recorded by the AWESOME system in Belgrade (Serbia) during solar
flares in the period between 10:40 UT and 13:00 UT on  April 22,
2011.

\end{summary}


\resthead{Influence of solar spectral lines on electron
concentration in terrestrial ionosphere  } {A. Nina, V. {\v C}ade{\v
z}, V. Sre{\'c}kovi{\'c}, D. {\v S}uli{\'c}}

\sectionb{1}{INTRODUCTION}

Solar radiation has a dominant external influence on the sunlit
Earth's atmosphere. The rate of photo-ionization processes in the
ionosphere depends on composite particles as well as on the solar
radiation spectrum at the considered altitude. The data on some
solar spectral lines could be of significant importance in studies
of solar flares (see Valn{\'{\i}}{\v c}ek \& Ranzinger (1972) and
references therein). For example, Valn{\'{\i}}{\v c}ek \& Ranzinger
(1972) studied the observed widths of the  H${\alpha}$ line for
different types of flares.

During the transmission through the terrestrial atmosphere, the
solar radiation is being attenuated and, at the altitude of about 70
km, only X rays and Lyman $\alpha$ line remain important i.e. with
sufficient energy to induce a noticeable photo-ionization of
particles in the ionospheric D-region.  The main subject of this
paper is a study of the intensity of these lines and comparison of
their time dependences in cases of unperturbed and perturbed
ionosphere during different solar flares with the corresponding
electron concentrations.

The electron concentration of plasma was calculated by using the
developed and elaborated method in Grubor et al. (2008), Zigman et
al. (2007) and Sulic et al. (2010). This calculations are based on
electron concentration influence on propagation properties of very
low frequency (VLF) waves. We used the time dependent solar spectrum
during solar flares according to the model presented in Chamberlin
et al. (2008).

\sectionb{2}{THEORY AND RESULTS}

Within the time interval 10:40 - 13:00 UT on April 22, 2011, the
DHO signal (transmitted from Germany at frequency 23.4 kHz) recorded at
Belgrade station has two dominant amplitude and phase increases.
The measurable features of the VLF signal refers to the change in
amplitude $\Delta A_{rec}$, measured in dB, relative to the ambient levels
prior to the event. The associated phase change $\Delta P_{rec}$ is
measured too (Figure 1).

\begin{figure}[!tH]
\vbox{
\centerline{\psfig{figure=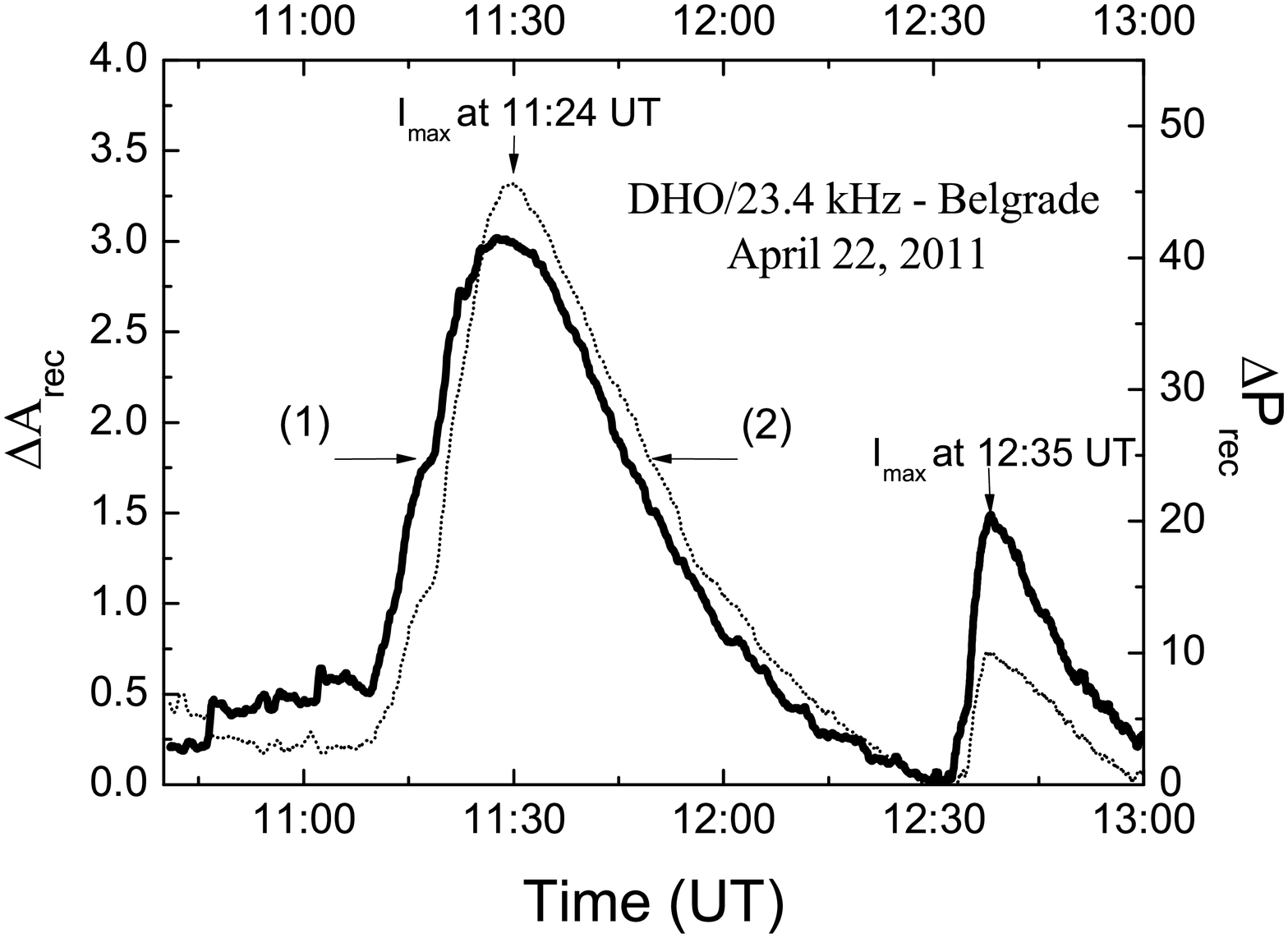,width=86mm,angle=0,clip=}}
\vspace{1mm} \captionb{1} {Perturbed amplitude (1) and phase (2) of
a signal emitted from DHO transmitter (Germany) and recorded by the
AWESOME receiver in Belgrade (Serbia) during the observed flares.
Null values correspond to the amplitude and phase recorded for the
unperturbed ionosphere.} }
\end{figure}

To calculate the electron concentration $N(t,h)$ in the D-region at
altitude $h$, we used the equation (Wait 1964):
\begin{equation}
N(t,h) =
1.43\cdot10^{13}e^{-0.15H'(t)}e^{(\beta(t)-0.15)(h-H'(t))}~.
\end{equation}
The parameters of the Wait's model of ionosphere, the reflection
height $H'(t)$ and  sharpness $\beta(t)$, are calculated by the LWPC
computer program (Ferguson 1998). The time distributions of the
electron concentration at four altitudes, where the dominant
electron gain and electron loss processes are the photo-ionization
and recombination, respectively, are presented in Figure 2. One can
see that at higher altitudes the photo-ionization processes strongly
dominate the recombination ones. Enhancements of $N(t,h)$ result
from increased solar radiation but the question remains which
particular lines in solar irradiance spectrum are important for the
photo-ionization processes in D-region. In introduction, we said
that these can be the lines from the X spectrum and Lyman $\alpha$
line. The time distribution of these lines is presented in Figure 3.
Here we can see that two X-flares and two increases of the Lyman
$\alpha$ line intensity (121.5 nm curve) occurred within the
observed time interval. Comparing Figure 2 with Figure 3 we can
notice that the peaks in the time distribution of the electron
concentration follow the time variation pattern of the X-emission
spectrum but not that of the Lyman $\alpha$ line intensity. This
leads to a conclusion that the influence of the X-spectrum lines is
dominant for electron concentration at altitudes of about 70 km. Two
intensity maximuma of all considered X-spectrum lines occur at 11:24
UT and 12:35 UT, which is 5 min before the time the corresponding
maxima appear in the electron concentration curve.


\begin{figure}[!tH]
\vbox{
\centerline{\psfig{figure=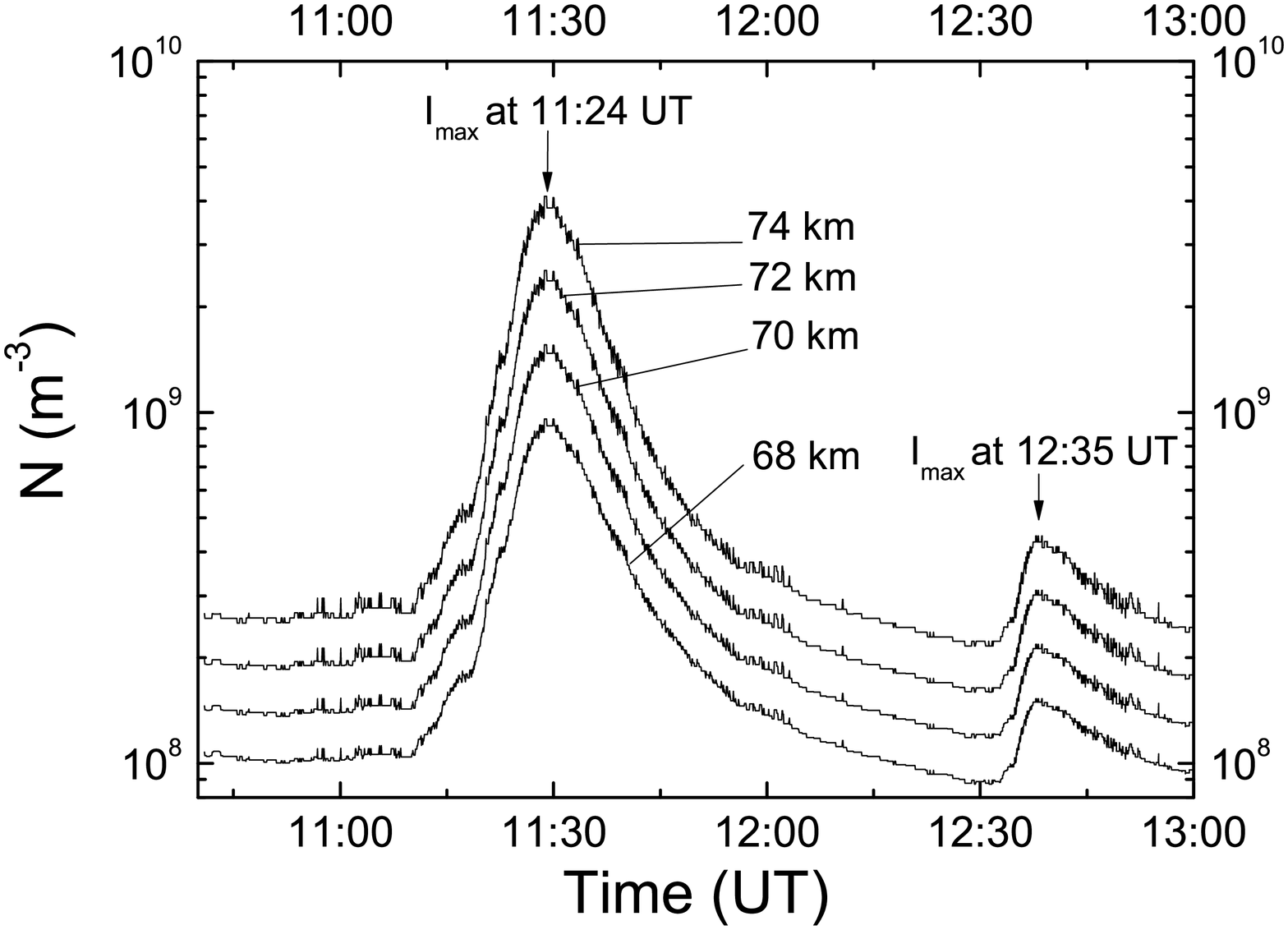,width=86mm,angle=0,clip=}}
\vspace{1mm} \captionb{2} {Electron concentrations on different
altitudes related to two flares of classes C7.7 and C3.0.} }
\end{figure}

\begin{figure}[!tH]
\vbox{
\centerline{\psfig{figure=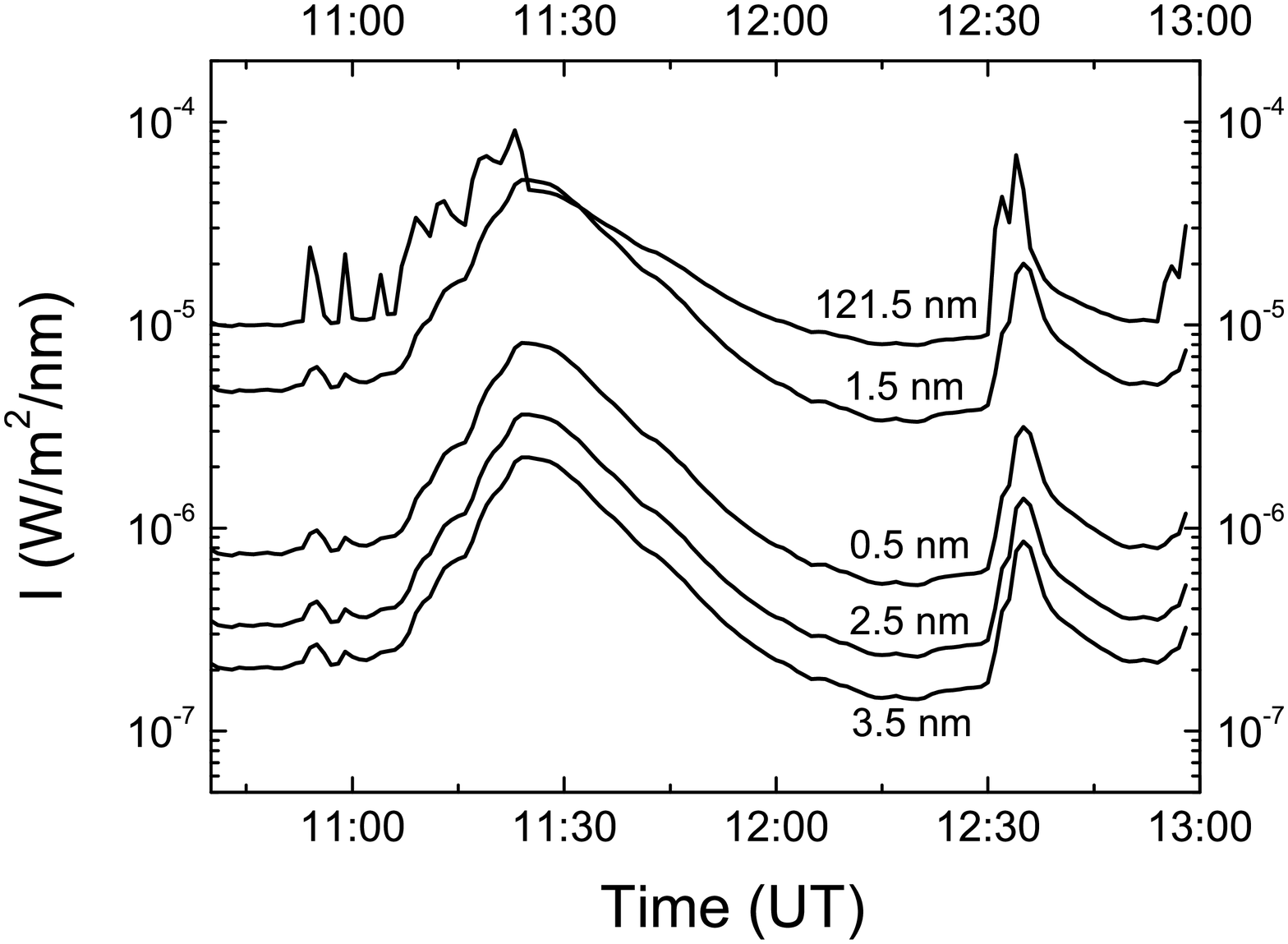,width=86mm,angle=0,clip=}}
\vspace{1mm} \captionb{3} {Time distribution of the solar radiation
intensity $I$ which has the dominant role in perturbing the
ionospheric D-region.} }
\end{figure}

The spectrum of solar radiation below 150 nm, presented in Figure 4,
shows that the analyzed lines in Figure 3 exhibit a bigger rise of intensity
then the other lines in the time interval of two maximal X ray
intensities occurring at 11:24 UT and 12:35 UT. Also, we can see that the first
X-flare produces higher line intensities which  followed by a higher
rise of electron concentration.

\begin{figure}[!tH]
\vbox{
\centerline{\psfig{figure=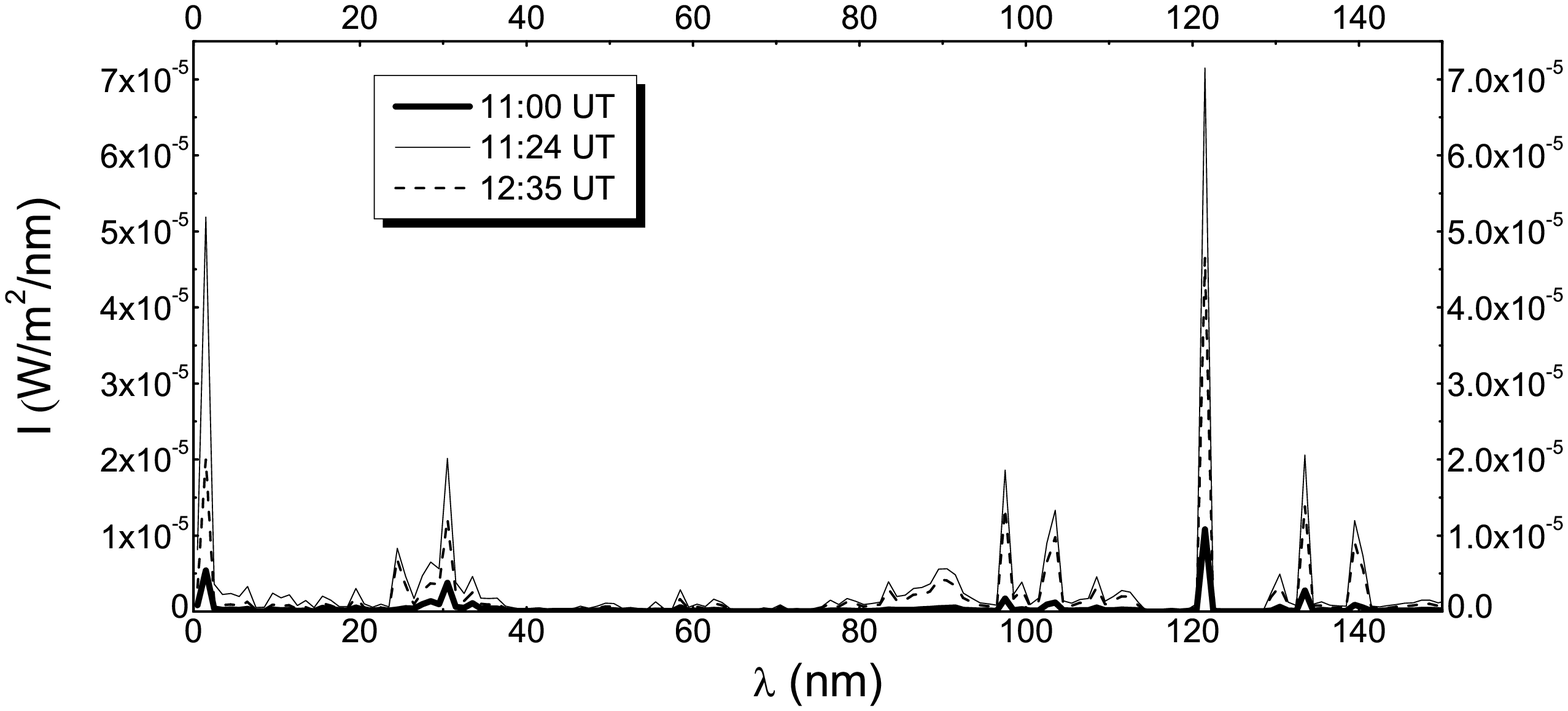,width=135mm,angle=0,clip=}}
\vspace{1mm} \captionb{4} {Spectrum of solar radiation for the
unperturbed ionosphere (dot line), at the maximum of the first (dash
line), and at the maximum of the second (solid line) flare.} }
\end{figure}

\sectionb{3}{CONCLUSIONS}

The analyzes presented in this work yield a conclusion
that the dominant influence on enhancement of the electron concentration
in the D-region during the observed solar flares is due to the increased
intensity of radiation lines in the X-spectrum, i.e.
 X-flares with bigger line
intensities produce larger rises of electron concentration.
This conclusion follows from the fact that the rate of the photo-ionization process
is higher then that of the recombination
processes within the time interval between the start of the irradiance increase and
to a few minutes after the irradiance maximum. In the case of these two flares,
the maximum of electron concentration occurred after the appearance of the maximum intensity
of lines produced by the X-flares.

\thanks{ The authors are thankful to the Ministry of Education and
Science of the Republic of Serbia (project numbers III 44002, 176002, 176004).}

\References

\refb Chamberlin P. C., Woods T. N., Eparvier F. G. 2008, Space
Weather, 6, S05001

\refb Ferguson J.A. 1998, Space and Naval Warfare Systems Center,
San Diego, CA

\refb Grubor D., {\v S}uli{\' c} D., {\v Z}igman V. 2008, Annales
Geophysicae, 26, 1731

\refb {\v S}uli{\' c} D., Nina A., Sre{\' c}kovi{\' c} V. 2010,
Publications of the Astronomical Observatory of Belgrade, 89, 391

\refb Valn{\'{\i}}{\v c}ek B., Ranzinger P. 1972, Bulletin of the
Astronomical Institute of Czechoslovakia, 23, 318

\refb Wait J.R., Spies K.P. 1964, National Bureau of Standards
Technical Note, 300

\refb {\v Z}igman V., Grubor D., {\v S}uli{\' c} D. 2008, Journal of
Atmospheric and Solar-Terrestrial Physics, 69, 775

%

\end{document}